\documentclass[aps,prl,floatfix,twocolumn]{revtex4-1}
\usepackage{times}
\usepackage{graphicx}% Include figure files
\usepackage{amsmath}
\usepackage{amssymb}
\usepackage{textcomp} %straight µ

\begin{document}
\title{Nanoscale confinement of energy deposition in glass by double ultrafast Bessel pulses}

\author{Jesus del Hoyo$^{1,2}$,Remi Meyer$^{1}$,Luca Furfaro$^{1}$,Francois Courvoisier$^{1,\ast}$\\
$^{1}$ FEMTO-ST institute, Univ. Bourgogne Franche-Comt\'e, CNRS,\\
15B avenue des Montboucons, 25030, Besan\c{c}on Cedex, France\\
$^{2}$ Applied Optics Complutense Group, Optics Department, \\
Universidad Complutense de Madrid, Facultad de Ciencias Fisicas, \\
Plaza de las Ciencias, 1, Madrid, 28040, Spain \\
$^{\ast}$ Corresponding author francois.courvoisier@femto-st.fr\\
\vspace{2cm}
This is a post-peer-review, pre-copyedit version of an article published in Nanophotonics (De Gruyter). The final authenticated version is available online at: \\
https://doi.org/10.1515/nanoph-2020-0457
\vspace{2cm}}

% Force line breaks with \\

%\title{Nanoscale confinement of energy deposition in glass by double ultrafast Bessel pulses}
%\author{Jesus del Hoyo,Remi Meyer,Luca Furfaro,Francois Courvoisier}

%\email{francois.courvoisier@femto-st.fr}
%\affiliation{FEMTO-ST institute, Univ. Bourgogne Franche-Comt\'e, CNRS, 15B avenue des Montboucons, 25030, Besan\c con Cedex, France}%
%\affiliation[2]{Applied Optics Complutense Group, Optics Department, Universidad Complutense de Madrid, Facultad de Ciencias Físicas, Plaza de las Ciencias, 1, Madrid, 28040, Spain}%

%\runningtitle{[Nanoscale confinement of energy deposition...]}

\begin{abstract}
Ultrafast laser pulses spatially shaped as Bessel beams in dielectrics create high aspect ratio plasma channels whose relaxation can lead to the formation of nanochannels. We report a strong enhancement of the nanochannel drilling efficiency with illumination by double pulses separated by a delay between 10 to 500~ps. This enables the formation of nanochannels with diameters down to 100~nm. Experimental absorption measurements demonstrate that the increase of drilling efficiency is due to an increase of the confinement of the energy deposition. Nanochannel formation corresponds to a drastic change in absorption of the second pulse demonstrating the occurrence of a phase change produced by the first pulse. This creates a highly absorbing long-living state. Our measurements show that it is compatible with the semi-metallization of warm dense glass which takes place within a timescale of $<$10~ps after the first laser pulse illumination.
\end{abstract}
\maketitle

%%%%%%%%%%%%%%%%%%%%%%%%%%%%%%%%%%%%%%%%%%%%%%%%%%%%%%%%%
%%%%%%%%%%%%%%%%%%%%%%%%%%%%%%%%%%%%%%%%%%%%%%%%%%%%%%%%%%

\section{Introduction}
Transparent dielectrics are ubiquitous in modern technology, but their structuring at the nanometric scale is a difficult task.  The main advantage of ultrafast laser-based processing techniques is the capability of structuring the material in three dimensions as well as processing elongated straight \cite{Bhuyan2010} or curved \cite{Froehly_2011} voids in single shot. This is possible because of the nonlinearity of the energy deposition process. It allows for a very precise control and versatility of the manufacturing process. 

 \begin{figure*}
    %\centering
    \includegraphics[width = 0.9\textwidth]{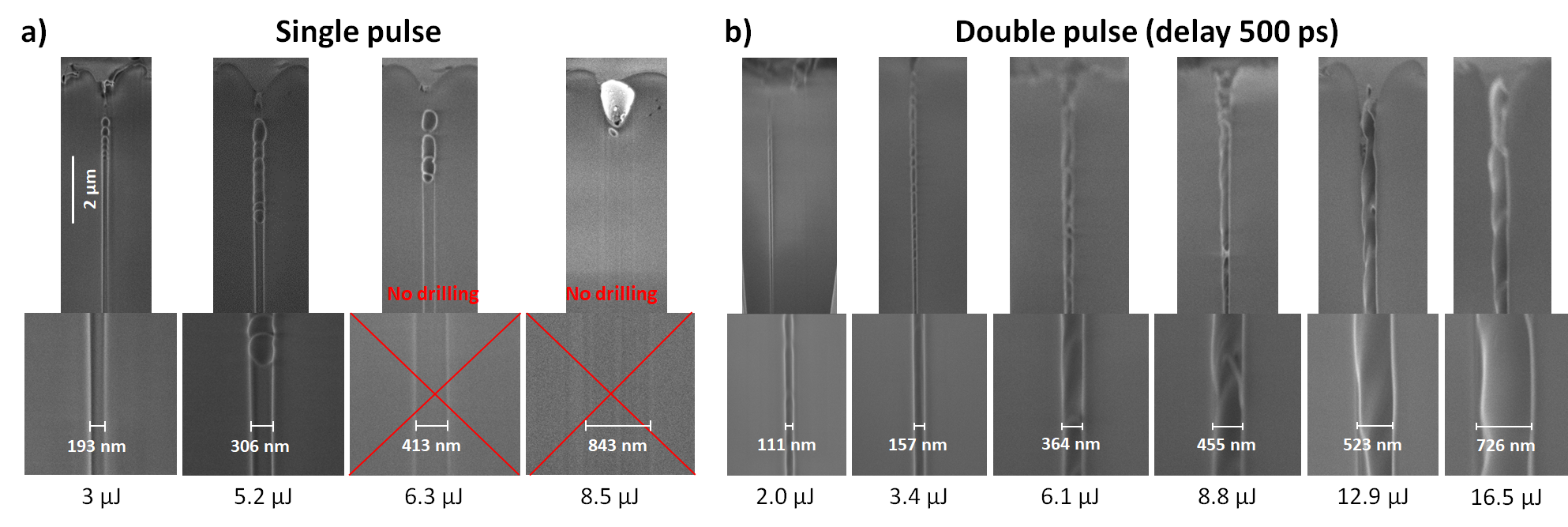}
    \caption{SEM images of channels drilled using single pulses (a) and double pulses (b). The insets correspond to a zoom in the images. The diameter of the channels is shown in the insets. The energy indicated corresponds to the total energy.}
    \label{fig:SEM}
\end{figure*}

The formation of high aspect ratio voids with diameters in the 100's nanometers range using only a single ultrafast laser pulse is particularly useful for the creation of nanophotonic crystals and Bragg gratings\cite{Bhuyan2010,Mikutis2013} or for glass cutting, in the framework of so called "stealth-dicing" technique  \cite{Ahmed_2008,Tsai2013,Bhuyan2015,Mishchik2016b,Meyer2017,Jenne2018,Meyer2019} which enables cutting glass using laser illumination speeds exceeding tens of centimetres per second.

Drilling of high aspect ratio nanochannels is made particularly controllable when using zeroth-order Bessel beams \cite{Durnin1987, Bhuyan2010,Alexeev2010,Boucher2018,Jenne2018,Meyer2019}. for sufficiently high focusing angles, Kerr effect is negligible and energy absorption occurs mainly in the central lobe.  
Although the mechanism leading to void channel formation in transparent dielectrics after the energy deposition stage is still an open question,  \cite{Glezer1997,Vailionis2011,Bhuyan2017}, our problematic here is that depending on the material and illumination geometry, limitations arise on the maximal and minimal nanochannel diameters that can be processed with Bessel beams. This is exemplified in Fig.\ref{fig:SEM}(a). It shows Scanning Electron Microscopy (SEM) images of nanochannels processed in Schott D263 glass with single femtosecond pulses. The pulse energy range over which a nanochannel can be opened is very limited. Increasing the pulse energy did not lead to an increase of the channel diameter above $\sim$~300~nm, in contrast with a relatively similar glass, Corning 0211 where nanochannels could be processed with diameters from 200 to 800~nm \cite{Bhuyan2010} with a twice shorter Bessel beam length.  

Here, we investigate how double pulses can solve this issue. Surface ablation by double Gaussian pulses conventionally sees a decrease of the ablation efficiency. This has mainly been attributed to a reduction of nonlinear ionization due to the intensity decrease when splitting the pulse in two, as well as to a screening effect from the plasma \cite{Stoian2002,Semerok2004,Chowdhury2005,Cao2018,Gaudfrin2020}.  
In contrast, the case of in-volume material excitation, at fluences lower than the ablation threshold, shows stronger modifications of fused silica (index of refraction, nanogratings formation) with double pulses in comparison with single pulses of identical total energy, for inter-pulse delay typically on the order of 10-100~ps \cite{Nagata2005,Wortmann2007,Chu2017,Wang2017,Wang2018,Stankevic2020}. The stronger modifications were attributed to an increase of the total deposited energy because of a higher absorption of the second pulse by the material softened after the first pulse. The first pulse generates a plasma of free-electrons and holes or ions that decays into self-trapped excitons and color centers. Those defects increase the absorption efficiency of the second pulse.

 In this article, we report that splitting the pulse energy in two equal energy pulses allows for drastically increasing the nanochannel formation efficiency. Nanochannels with $\sim$~100~nm diameters can be reliably processed, and we can increase by a factor two the maximal channel diameter compared to single shot case. We have studied the evolution of the channel morphology with inter-pulse delay and have characterized the evolution of absorption as a function of delay and energy. The overall absorption is slightly less for the double pulse case than for the single pulse one. But our results demonstrate for the first time that pulse splitting with a delay in the 100-500~ps allows for enhancing the {\it confinement} of energy deposition.
 
 An analysis of the evolution of the second pulse absorption in time reveals two different dynamics respectively for sub- and above threshold for nanochannel drilling. While at the lowest energies, the absorption dynamics follows the decrease conventionally expected by the formation of self-trapped excitons, our results at higher energies,those which allow nanochannel formation, show a striking new behaviour for bulk excitation. These are compatible with the generation of warm dense silica within $\sim$ 10~ps after the first pulse, which has a very localized and high absorptivity because of  semi-metallization. The Bessel beam geometry associated to a pulse burst in the GHz repetition rate regime \cite{Kerse2016} is therefore expected to lead to highly efficient formation of Warm Dense Matter with tabletop experiments, which is important for the fundamental understanding of astrophysical objects \cite{Bonitz2020}.

\section{Experimental setup}

\begin{figure}
    \centering
    \includegraphics[width = \columnwidth]{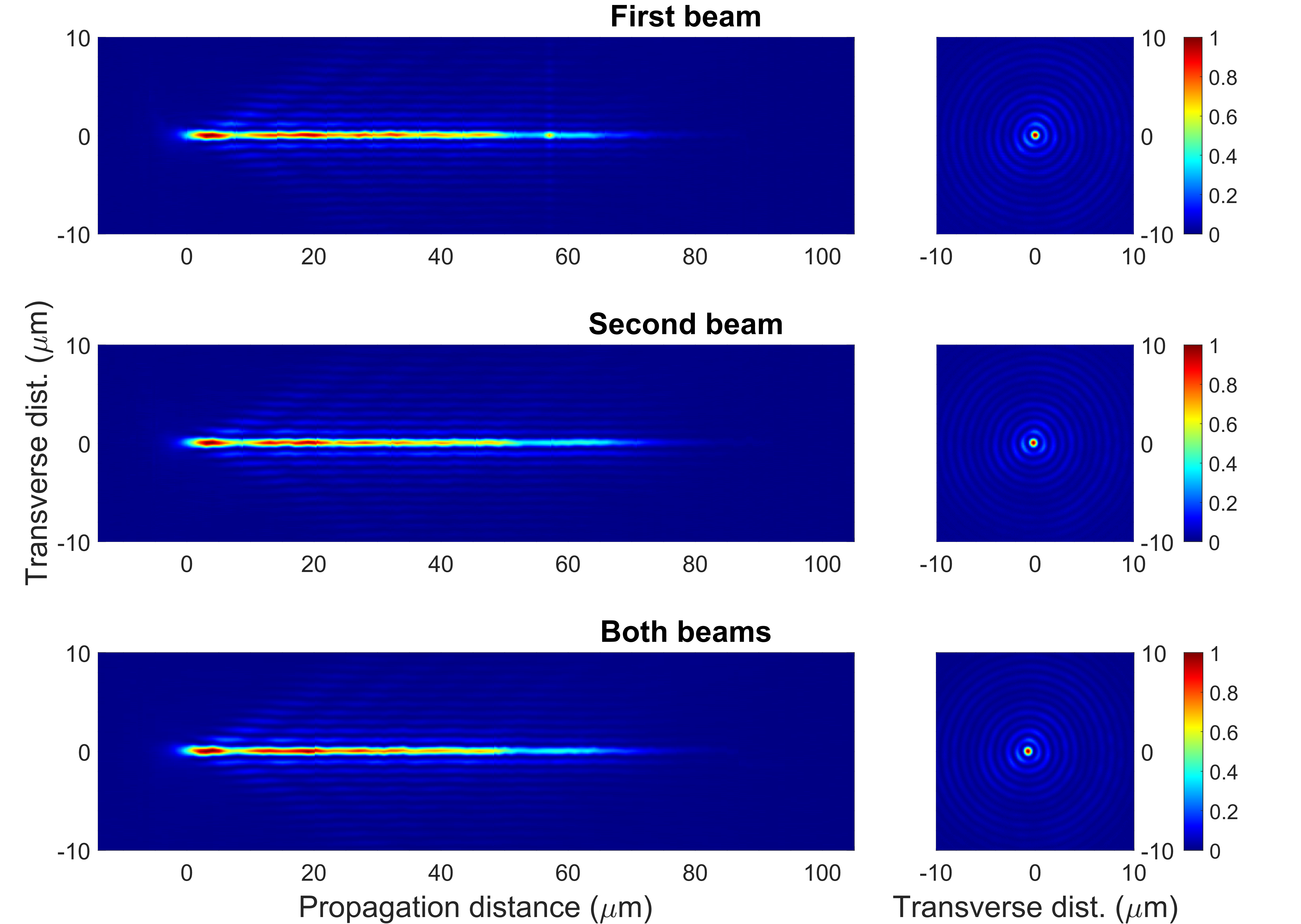}
    \caption{(Left column): Experimental longitudinal intensity distribution for single and double pulses.  (Right column): Corresponding experimental intensity cross sections. All images are normalized to their own maximum. }
    \label{fig:beams}
\end{figure}

 Our experimental setup is based on a 5~kHz ultrafast Ti-sapphire laser source with 800~nm central wavelength. Before beam shaping, the ultrafast pulses were split into equal parts and recombined using a Mach-Zehnder interferometer arrangement. The pulses were independently circularly polarized with the same rotation orientation. The inter-pulse delay is controlled using a micrometric motorized translation stage and can be varied up to a maximum of 500~ps. The zero-delay was characterized using the interference between the pulses. Beam shaping was performed using an axicon with base angle 1$^{\circ}$ and a telescopic arrangement with magnification $\sim$110 to generate a Bessel beam with a length of $\sim$100 $\mu$m and a cone angle of 26$^{\circ}$ in air.

A computer-controlled Pockels cell was used for single shot illumination. To avoid any potential cumulative effects due to the irradiation by the small polarization leakage during the dead time between the sample positioning to a fresh area and the effective illumination by the high intensity pulses, two additional fast mechanical shutters were installed with effective opening time of $\sim$2~ms. In practice, we have observed that the presence of these fast shutters had no effect in the irradiations, ensuring that the leaks from the Pockels cell are absolutely negligible.

 The pulse energy and duration were characterized at the sample site. The pulse duration is 100~$\pm$~15~fs. We used recently cleaved Schott D236 borosilicate glass slides as sample. The sample flatness was ensured to be below 0.1~mrad.
%%%
 Beam characterization was performed using a microscope objective with numerical aperture 0.8, which was combined to a lens to form a telescope with a magnification factor of 56 \cite{Froehly2014}. Figure \ref{fig:beams} shows the intensity distribution of the generated beams. We highlight the fact that the high magnification ($\times$110) used for beam shaping also magnifies any angular deviation from a misalignment in the Mach-Zehnder interferometer. Even the small flatness deviation of the high-precision translation stage generates a deviation by more than 1~$\mu$rad when translating the mirrors to change the inter-pulse delay from 0 to 100~ps. This deviation is enough to prevent the accurate superposition of the central lobes of the two Bessel beams. The bottom panel in Fig. \ref{fig:beams} shows the perfect beam superposition, which was corrected for each inter-pulse delay.

The samples were characterized first using conventional optical microscopy with through-illumination using a microscope objective equipped with a variable correction collar. This was followed by Focused Ion Beam (FIB) milling where a careful procedure was developed to avoid modifying the laser processed structures as in reference \cite{Rapp2016}. This allows side-imaging the nanochannels by scanning electron microscopy (SEM).

In a second step, we conducted absorption measurements of the single pump pulse or of the double pump one. Pulse absorption was characterized using the ratio of the signals between two photodiodes positioned before and after the sample. The photodiodes were large photodiodes (100 mm$^2$) to ensure the whole beam was collected. 
The reference photodiode was placed after a 10\% splitter before the first telescope. After the sample, the beam was collimated by a microscope objective with numerical aperture 0.8 and the far-field was imaged onto the second photodiode. The acquisition was performed in single shot using an oscilloscope. The linearity of the acquisition was specifically ensured over the whole energy range investigated. For the absorption measurements, the Bessel beam was fully enclosed within the sample thickness, and the sample was also translated so that each shot illuminates an undamaged area. 

 Because of the Bessel beam conical structure, reflection and transmission on the plasma are indistinguishable. Both transmitted and reflected beams are directed in the forward direction with the same angle with respect to the optical axis. The overall absorption $A$ is therefore  complementary to the measured transmission: $A=1-E / E_0$, where $E$ and $E_0$ are the energies collected after the sample with the Bessel beam respectively placed inside and outside of the glass sample. Therefore, the Fresnel losses are eliminated from the absorption curves. The experiment was repeated 50 times with the same parameters and the results were averaged.

%%% Experimental results
\section{Results}
Figure \ref{fig:SEM}(a) shows SEM images of Schott D263 glass samples after single pulse irradiation. We observe that channels of diameters $\sim$ 200 and 300~nm are drilled for energies of respectively 3 and 5.2~$\mu$J. However, further increase of the pulse energy does not lead to channel opening but instead to a density modification. This modification extends over a diameter which increases with pulse energy. We infer that at the highest energies shown here, a plasma is formed over a too wide diameter, which limits the thermodynamical gradients needed to generate microexplosions and void formation.

%%%
We repeated the same experiment for double pulses with inter-pulse delay of 500~ps and the results are shown in  Fig. \ref{fig:SEM}(b). We readily observe that the morphology of the channels is relatively similar to the single pulse case, except that, now, nanochannels are opened in all cases. The produced channels width extends from $\sim$100 to more than 700~nm. This is a much wider processing window than in the single pulse case. The small inhomogeneities along the channels are attributed to the intensity inhomogeneities of the beam as can be seen from Figure \ref{fig:beams} \cite{Dudutis2018}.

\begin{figure}
    \centering
    \includegraphics[width = \columnwidth]{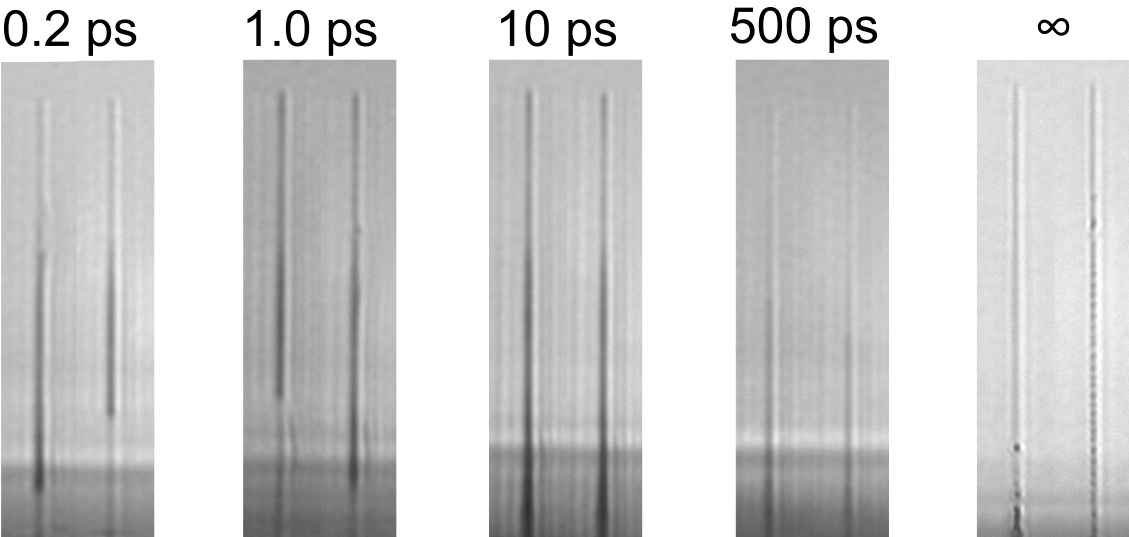}
    \caption{Optical microscopy images of several channels produced using double pulses for a total energy of 5~$\mu$J. Two channels in each image were obtained with the same parameters. The images of each channel have been cropped and stitched together to save figure space.}
    \label{fig:Variation_withDelay}
\end{figure}

We will now investigate the influence of the inter-pulse delay. For this, we show in Fig. \ref{fig:Variation_withDelay} images of nanochannels under optical microscopy with five inter-pulse delays. For each case, the drilling experiment has been repeated several times, so as to evaluate the repeatability of the laser-induced material modification. In the figure we show two images of channels obtained in identical conditions. We note that the distance between two channels is 20~$\mu$m to avoid any influence of a material modification by a previous pulse on the next modifications. Under our optical characterization system, voids with diameter $\gtrapprox$~400~nm appear deep black.

Delays below the pulse duration produce results that vary very significantly from shot to shot (not shown). A possible explanation is that the phase between the two pulses cannot be controlled with sufficient accuracy to maintain the same interference state from shot to shot. Strong shot to shot variations were similarly observed in the absorption experiments as described below.

For inter-pulse delays from 1~ps and up to 500~ps, irradiations produced channels over the full tested range of energies (2 - 16.5~$\mu$J total energy), in high contrast with the single pulse case. This is confirmed from SEM measurements. For inter-pulse delays between 0.2 to $\sim$~10~ps, a transition occurs where the channels appear wider, more uniform along their length and more reproducible on a shot to shot basis.  For delays of 10~ps and above, the structures show a higher degree of uniformity and repeatability. We noted that the optimum depends on the total pulse energy. In the case where the inter-pulse delay was of several tens of ms, {\it i.e.} when the excitation by the first pulse has fully relaxed before the second pulse arrives (noted as $\infty$ in the figure), the structures are much more faint and less regular. This indicates the state reached at 500 ps is not identical to state reached after the complete relaxation. 

Noticeably, the marks of thermal effects are highly reduced on the optical microscopy images and on the SEM images of Fig. \ref{fig:SEM}(b) in comparison with what is conventionally observed with picosecond illumination in glass \cite{Goette2016} or in sapphire \cite{Rapp2016}, or in comparison with illumination via a second pulse with nanosecond duration or more \cite{Lin2010,Ito2018}. We infer that the thermodynamical pathways (ionization dynamics, pressure, temperature...)  are different between the case of double femtosecond pulses excitation and the case of illumination by picosecond pulses.

\begin{figure}
    %\centering
    \includegraphics[width = 0.95\columnwidth]{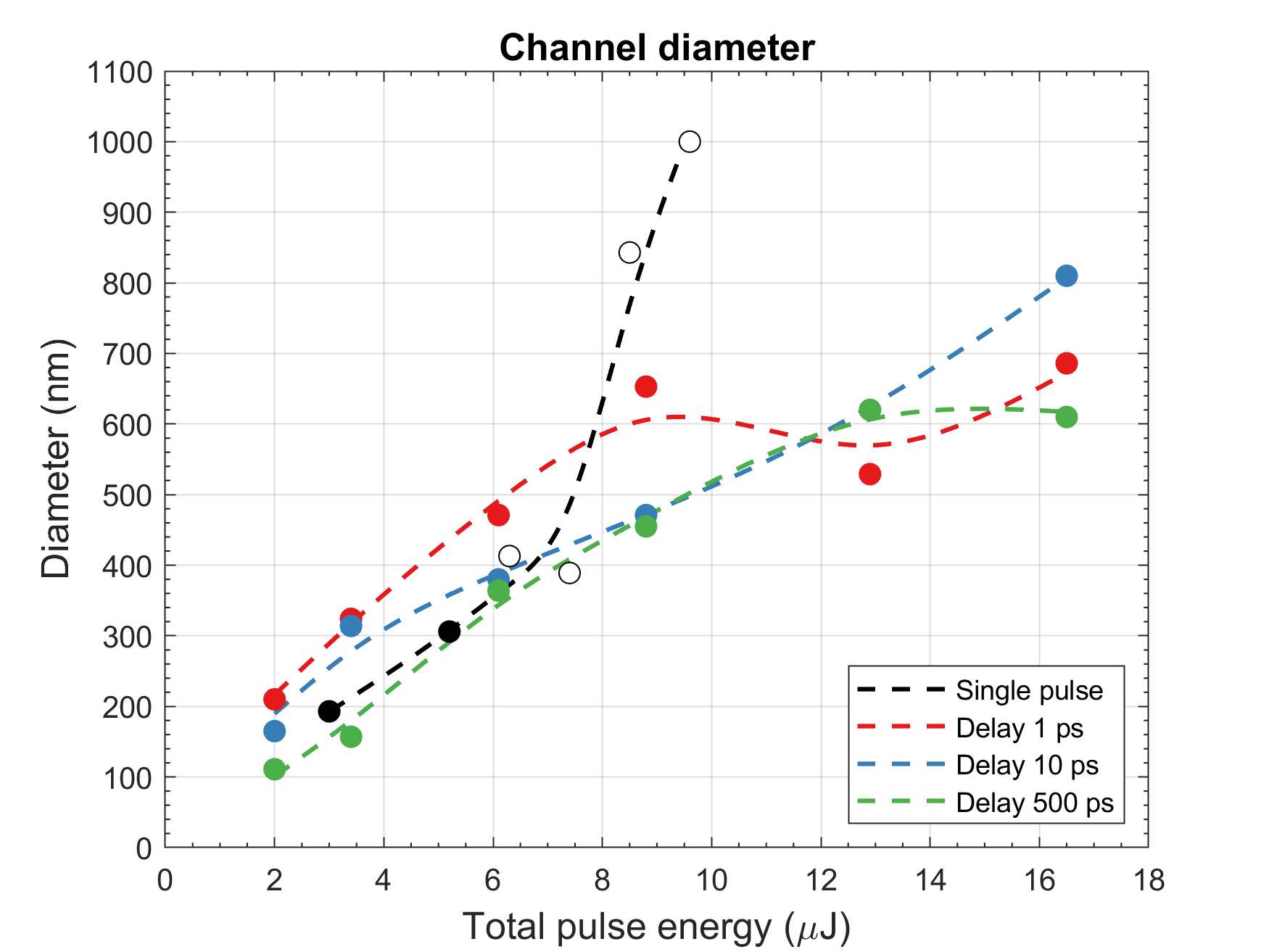}
    \caption{Channel diameter measured on SEM images for modifications made using single and double pulses with inter-pulse delays of 1, 10 and 500 ps, at a depth of 5~$\mu$m below the surface of the output facet. Filled dots correspond to the formation of a void channel, while circles correspond to refractive index modification. Dashed lines correspond to B-spline interpolation. The horizontal axis corresponds to the total energy in all cases.}
    \label{fig:diameter_vsE}
\end{figure}

Figure \ref{fig:diameter_vsE} quantitatively compares the evolution of the channel diameter with the total input pulse energy for single pulses and three delays. The measurements have been performed using SEM images of the channel cross-sections, after FIB milling.  
The black markers show the evolution of the diameter of the modification produced in the single shot regime. The error bar on the measurement is due to the accuracy of FIB milling between two SEM images and is $\sim$40~nm. Voids are shown with black disks, density modifications with the circles. The three other lines show the results for inter-pulse delays of 1, 10 and 500~ps. We first notice that for the three cases in double pulse regime, the evolution is approximately the same, with a quasi-linear increase with pulse energy. This follows the same trend as shown originally in reference \cite{Bhuyan2010}. At the lowest energies, the longest delays produce the smallest diameters. This variation can be attributed to a decay of the material excitation by the first pulse. For a delay of 1~ps, at energies above 8~$\mu$J, we observe a fluctuation of the diameter. This variation can be attributed to the strong phase transition that we describe in the next section, which occurs over a scale of 1-10~ps depending on the deposited energy.

The differences between single and double pulses is noticeable. First, double pulses allow reaching extremely small diameters down the $\sim$100~nm,{\it i.e.} nearly eight times lower than the laser wavelength. For the glass used here (Schott D263), the maximal diameter reached with double pulses is almost three times higher than in the single pulse case. These results suggest that wider channels could be produced using higher pulse energies. We expect that the maximal channel diameter achievable will depend on the inter-pulse delay.

\begin{figure}
    \centering
    \includegraphics[width = 0.9\columnwidth]{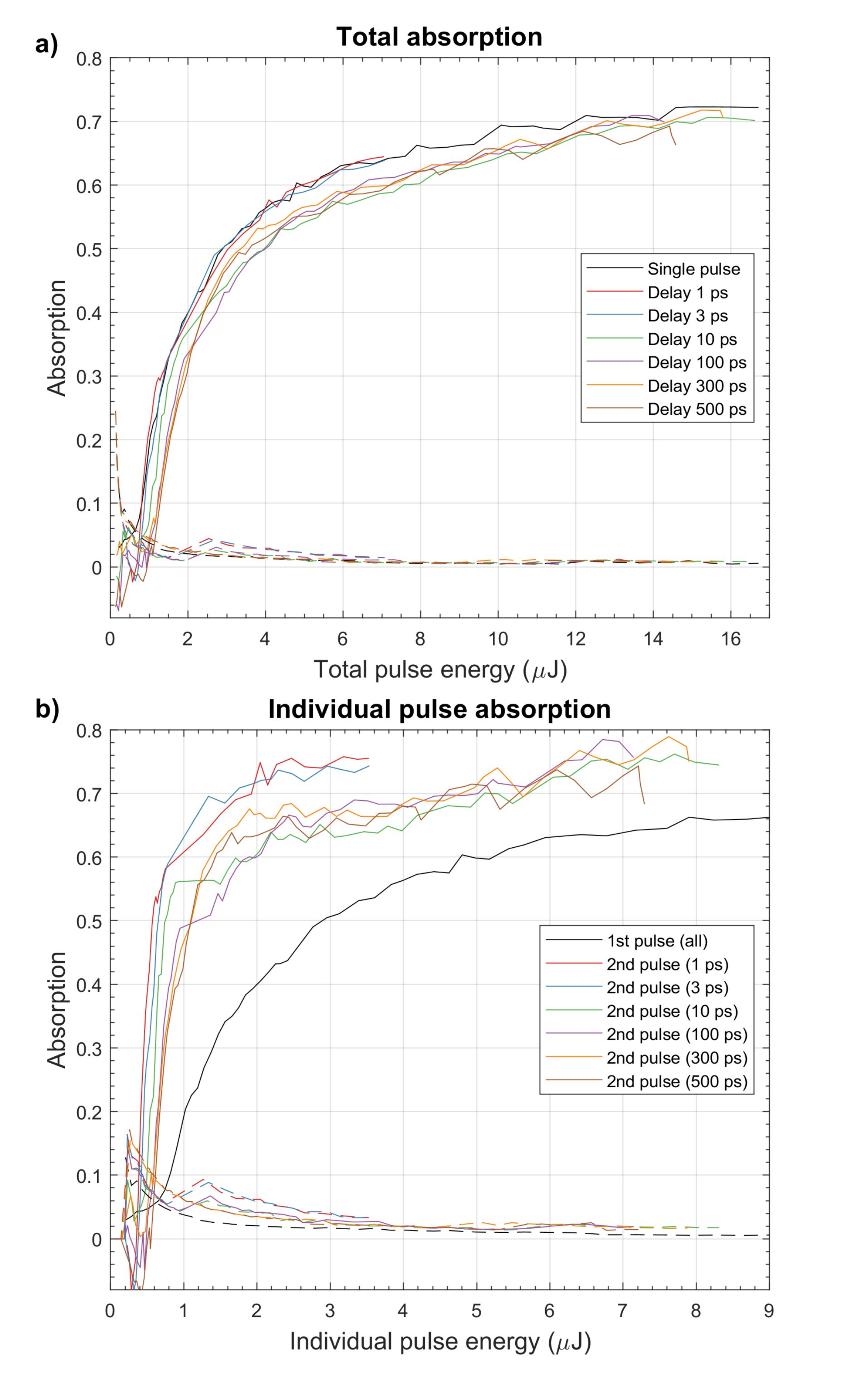}
    \caption{Absorption measurements of single and double pulses. (a) Total absorption of double pulses. (b) Absorption of the first and the second pulse. In (a) and (b), the dashed lines show the value of the standard deviation over 100 shots.}
    \label{fig:Absorption}
\end{figure}

\begin{figure}
    \centering
    \includegraphics[width= 0.9\columnwidth]{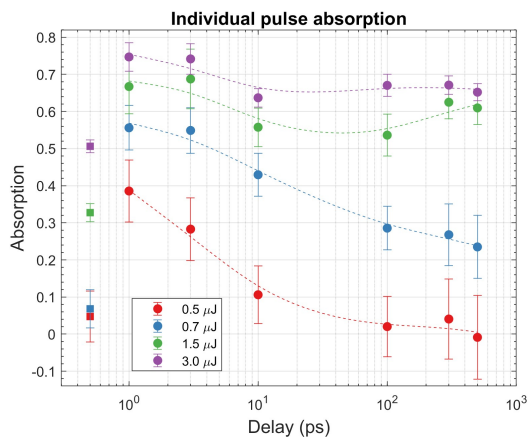}
    \caption{Second pulse absorption as a function of the inter-pulse delay, for different {\it individual} pulse energies. For reference, we indicate with square markers the absorption of the first pulse. Dashed lines correspond to B-spline interpolation. The error bars correspond to one standard deviation.}
    \label{fig:AbsDelay}
\end{figure}

\section{Pulse absorption}

A priori, two origins can explain a higher drilling efficiency with double pulses than with single ones. This can be due to a higher overall absorption of energy or to the absorption of a possibly smaller amount of energy but on a significantly smaller diameter (the length of the modifications remains constant between single and double pulse illumination). We demonstrate here that it is the second case.

Figure \ref{fig:Absorption}(a) shows the total absorption for single and double pulses as a function of the total input pulse energy.
It can be seen that the absorption is overall similar in all cases, thus, splitting the pulse does not lead to a higher absorption. The absorption is in fact higher for the single pulse case (black curve), particularly for energies in the range 2-4~$\mu$J, where the slope is the highest and which also correspond to the threshold energy for channel drilling. Therefore, the highest drilling efficiency observed at 500~ps inter pulse delay does not correspond to a higher absorption. The higher absorption measured for the single pulse case can be attributed to a higher efficiency for the multiphoton and tunnel ionization processes. Those might take place either out of the central lobe of the Bessel beam or within this central lobe, but on a wider area than in the double pulse case. We conclude that the enhanced efficiency for void channel opening originates from a higher {\it density} of deposited energy.

The absorption of the second pulse can be straightforwardly derived from the overall absorption by subtracting the single pulse contribution.
Figure \ref{fig:Absorption}(b) shows the absorption of the first and second pulses, shown as a function of the {\it individual} pulse energy (we recall that first and second pulses have the same energy). 
Above an individual pulse energy of $\sim$~1~$\mu$J, the second pulse absorption drastically increases with a step-like curve, reaching absorption above 0.6, and well above the first pulse absorption at this energy. This is followed by a smoother increase, for energies greater that 2~$\mu$J. We note that the step in absorption occurs for an energy which nearly corresponds to the threshold for nanochannel drilling (2~$\mu$J total energy). The abrupt change in absorption shows that an important transition is crossed when the energy of the first pulse is sufficiently high. We will see below that this transition corresponds to the formation of warm dense matter.

To obtain a better view of the evolution of absorption in time, we plot in Fig. \ref{fig:AbsDelay} the evolution of the second pulse absorption as a function of the inter-pulse delay for a set of representative energies. Square markers show the first pulse absorption as a reference, for the same pulse energy.

For the lowest individual pulse energies, typically below 1~$\mu$J ({\it i.e} below threshold for nanochannel drilling in double pulse regime) the absorption effectively decreases with the inter-pulse delay (red and blue curves). This is consistent with the decay times of self-trapped excitons (STEs), which are on the order of 30 and 300~ps \cite{Grojo2010}. The higher absorption of the second pulse in comparison with the first can be attributed in this case to the formation of STEs which have a much higher ionization cross-section than glass because of their intermediate position in the bandgap. The evolution of the second pulse absorption with time at low pulse energies explain the dynamics observed by Wang {\it et al} using Bessel beams in fused silica in double pulse configuration: after chemical etching of the modifications inscribed, channel length highly increased for delays in the pulse separation window 1-50~ps and decreased for longer delays \cite{Wang2017}.

In contrast, at energies above the threshold for channel drilling ({it e.g.} green and purple curves for 1.5 and 3~$\mu$J), the absorption curves show only a small decrease with delay up to 10~ps and then stay nearly constant with time. We also note that the curves with different delays are nearly identical within the fluctuations in Fig. \ref{fig:Absorption}(b) above 1.5~$\mu$J.
This indicates that, in the case of nanochannel drilling, a transformation triggered by the first pulse occurs during 10~ps and reaches a state which has a typical lifetime exceeding the nanosecond scale. 

\section{Discussion}

Our observations are in high agreement with the results by Garcia-Lechuga {\it et al} of the measurements of the transient reflectivity and transmission of the surface of fused silica after irradiation by high fluence (7~J/cm$^2$) femtosecond pulses \cite{Garcia-Lechuga2017}. A state with absorptivity close to 1 and negligible reflectivity was characterized to form within the same scale of 1 to 10~ps. This state of matter was inferred by the authors to be a hot blackbody. In our case, the fluence reached for only 1~$\mu$J single pulse energy exceeds 10~J/cm$^2$ in the central lobe of the Bessel beam because of the extreme focusing angle (without accounting for potential nonlinear losses on the beam path). The fluence range in which we operate is therefore very similar to the conditions of the work mentioned above. We note that in our case, the pulse absorption cannot reach 1 because the plasma, with typical size of a few 100's nm, is smaller than the diameter of the Bessel beam central spot (660~nm FWHM, 1.2~$\mu$m in linear regime), so that the interaction cross section between the second pulse and the hot cylinder of matter is necessarily below 1. 
 
For an individual pulse energy of 1.5~$\mu$J, at the onset of the plateau of absorption in Figure \ref{fig:Absorption}(a), the energy density absorbed within the central lobe is already exceeding 8~MJ/kg, for a diameter of the excited material estimated to $\lesssim$~600~nm (order of magnitude of the central lobe FWHM of the Bessel beam). This corresponds to the typical energy density of Warm Dense Matter in fused silica \cite{Engelhorn2015}. Denoeud {\it et al} have experimentally determined with X-ray absorption near-edge spectroscopy that warm dense glass shows a semi-metallic behaviour\cite{Denoeud2014}. Engelhorn {\it et al} demonstrated that with similar energy density (10~MJ/kg) as in our experiment, warm dense silicon dioxide reaches an ionic temperature of 5~000~K and an electronic temperature on the order of 30~000~K within a duration of 20~ps \cite{Engelhorn2015}. Our results are fully in line with those, including the typical temporal scale of the transformation. 
 
Therefore, our results can be explained by the metallization of glass occurring within a couple tens of picosecond after the first pulse. Because of the drastic change in drilling efficiency observed between single and double pulses, absorption of the second pulse must occur within a significantly smaller volume than in the case of a single pulse. We interpret the increase of reproducibility as originating from a progressive increase of the confinement of absorption between 10 and 500~ps. The warm dense matter diameter shrinks while its absorption is expected to increase. The diameter shrinkage is associated to some energy loss, which makes, at the lowest energies, the single pulse slightly more efficient than the double pulse with 500 ps, but less efficient than with 1 and 10 ps delays (see Fig.\ref{fig:diameter_vsE}). We note that modelling the absorption properties of warm dense matter is for now computationally extremely challenging \cite{Bonitz2020}. The large variability observed at delays smaller than 10~ps is attributed to the phase change occurring during this time.

\section{Conclusion}

We have demonstrated that double pulse femtosecond illumination can drastically enhance the efficiency of nanochannel formation in comparison with the single pulse case. The morphology of the channels shows a negligible heat affected zone in contrast with conventional illumination with picosecond durations. Absorption measurements show that the increased void channel formation efficiency can be understood as a higher confinement of the energy deposition. At energies below the threshold for nanochannel opening, the absorption of the second pulse decreases with time as can be expected from the formation of self-trapped excitons. At higher energies, where nanochannel formation occurs, our experimental results indicate that the first pulse triggers a phase transition within 1-10~ps, and this state has a lifetime exceeding 1~ns. 

The higher efficiency for void formation is explained by the swift generation of the semi-metallic state of warm dense glass generated over a diameter significantly smaller than in the single pulse case, with a diameter typically lower than 600~nm. For all delays above 1~ps, channel formation was observed in double pulse illumination, and a smoother transition operates between 10 and several hundreds picoseconds. We note that our measurements cannot distinguish between the generation by the first pulse of a cylinder with a uniform density of warm dense matter or if a rarefied region at the center, surrounded by warm dense matter progressively forms at a scale of several 100's ps. Our results are therefore compatible with both expected mechanisms of nanochannel formation \cite{Vailionis2011, Bhuyan2017}.

Our results have implications both on applied and fundamental aspects. In the field of nano-machining with ultrafast laser pulses, the best configuration in the investigated delay range is for 10-500~ps depending on the energy. This time interval falls in the GHz repetition range. Our results can also explain the enhancement of localized energy deposition in glass using GHz repetition rate sources \cite{Kerse2016}. The multiple pulse configuration, at high total energies, can be used to increase the maximal void diameters, while at the lowest energies, fine tuning of the void diameter below $\sim$100~nm becomes possible. Further research with GHz sources is needed to evaluate an optimal number of pulses. Our results also demonstrate the capability of generating Warm Dense Matter within the bulk of materials by using burst of pulses. We believe a similar state can be created within other transparent materials such as sapphire. The confinement in the material's bulk using a nondiffracting Bessel beam increases the confinement in comparison with laser impacts on surface. This opens new routes to create and control high temperature and high pressure states within relatively high volumes, with length up to centimetre scale \cite{Meyer2019}, for tabletop laboratory scale experiments to reproduce astrophysical pressure and temperature conditions.

%%%%%%%%%%%%%%%%%%%%%%%%%%%%%%%%%%%%%%%%%%%%%%%%%%%%%%%%%%%
%%%%%%%%%%%%%%%%%%%%%%%%%%%%%%%%%%%%%%%%%%%%%%%%%%%%%%%%%%%
% If you have acknowledgments, this puts in the proper section head.
%\begin{acknowledgement}

We acknowledge the support and funding from  
European Research Council (ERC) 682032-PULSAR, Region Franche-Comte council,  the EIPHI Graduate School (ANR-17-EURE-0002), I-SITE BFC (ANR-15-IDEX-0003),  French RENATECH network.

Technical assistance by C. Billet, E. Dordor and fruitful discussions with R. Giust are gratefully acknowledged.
%\end{acknowledgement}

% Create the reference section using BibTeX:
%\bibliographystyle{ieeetr}
\bibliographystyle{naturemag}
\bibliography{Bibliography2}

\begin{thebibliography}{10}
\expandafter\ifx\csname url\endcsname\relax
  \def\url#1{\texttt{#1}}\fi
\expandafter\ifx\csname urlprefix\endcsname\relax\def\urlprefix{URL }\fi
\providecommand{\bibinfo}[2]{#2}
\providecommand{\eprint}[2][]{\url{#2}}

\bibitem{Bhuyan2010}
\bibinfo{author}{Bhuyan, M.~K.} \emph{et~al.}
\newblock \bibinfo{title}{High aspect ratio nanochannel machining using single
  shot femtosecond {Bessel} beams}.
\newblock \emph{\bibinfo{journal}{Appl. Phys. Lett.}}
  \textbf{\bibinfo{volume}{97}}, \bibinfo{pages}{081102}
  (\bibinfo{year}{2010}).

\bibitem{Froehly_2011}
\bibinfo{author}{Froehly, L.} \emph{et~al.}
\newblock \bibinfo{title}{Arbitrary accelerating micron-scale caustic beams in
  two and three dimensions}.
\newblock \emph{\bibinfo{journal}{Opt. Express}} \textbf{\bibinfo{volume}{19}},
  \bibinfo{pages}{16455} (\bibinfo{year}{2011}).

\bibitem{Mikutis2013}
\bibinfo{author}{Mikutis, M.}, \bibinfo{author}{Kudrius, T.},
  \bibinfo{author}{{\v{S}}lekys, G.}, \bibinfo{author}{Paipulas, D.} \&
  \bibinfo{author}{Juodkazis, S.}
\newblock \bibinfo{title}{High 90{\%} efficiency bragg gratings formed in fused
  silica by femtosecond {Gauss-Bessel} laser beams}.
\newblock \emph{\bibinfo{journal}{Opt. Mater. Express}}
  \textbf{\bibinfo{volume}{3}}, \bibinfo{pages}{1862} (\bibinfo{year}{2013}).

\bibitem{Ahmed_2008}
\bibinfo{author}{Ahmed, F.}, \bibinfo{author}{Lee, M.},
  \bibinfo{author}{Sekita, H.}, \bibinfo{author}{Sumiyoshi, T.} \&
  \bibinfo{author}{Kamata, M.}
\newblock \bibinfo{title}{Display glass cutting by femtosecond laser induced
  single shot periodic void array}.
\newblock \emph{\bibinfo{journal}{Appl. Phys. A}}
  \textbf{\bibinfo{volume}{93}}, \bibinfo{pages}{189--192}
  (\bibinfo{year}{2008}).

\bibitem{Tsai2013}
\bibinfo{author}{Tsai, W.-J.}, \bibinfo{author}{Gu, C.-J.},
  \bibinfo{author}{Cheng, C.-W.} \& \bibinfo{author}{Horng, J.-B.}
\newblock \bibinfo{title}{{Internal modification for cutting transparent glass
  using femtosecond Bessel beams}}.
\newblock \emph{\bibinfo{journal}{Opt. Eng.}} \textbf{\bibinfo{volume}{53}},
  \bibinfo{pages}{1--7} (\bibinfo{year}{2013}).

\bibitem{Bhuyan2015}
\bibinfo{author}{Bhuyan, M.~K.} \emph{et~al.}
\newblock \bibinfo{title}{High-speed laser-assisted cutting of strong
  transparent materials using picosecond {Bessel} beams}.
\newblock \emph{\bibinfo{journal}{Appl. Phys. A}}
  \textbf{\bibinfo{volume}{120}}, \bibinfo{pages}{443--446}
  (\bibinfo{year}{2015}).

\bibitem{Mishchik2016b}
\bibinfo{author}{Mishchik, K.} \emph{et~al.}
\newblock \bibinfo{title}{Ultrashort pulse laser cutting of glass by controlled
  fracture propagation}.
\newblock \emph{\bibinfo{journal}{J. Laser Micro/Nanoeng.}}
  \textbf{\bibinfo{volume}{11}}, \bibinfo{pages}{66--70}
  (\bibinfo{year}{2016}).

\bibitem{Meyer2017}
\bibinfo{author}{Meyer, R.}, \bibinfo{author}{Giust, R.},
  \bibinfo{author}{Jacquot, M.}, \bibinfo{author}{Dudley, J.~M.} \&
  \bibinfo{author}{Courvoisier, F.}
\newblock \bibinfo{title}{Submicron-quality cleaving of glass with elliptical
  ultrafast {Bessel} beams}.
\newblock \emph{\bibinfo{journal}{Appl. Phys. Lett.}}
  \textbf{\bibinfo{volume}{111}}, \bibinfo{pages}{231108}
  (\bibinfo{year}{2017}).

\bibitem{Jenne2018}
\bibinfo{author}{Jenne, M.} \emph{et~al.}
\newblock \bibinfo{title}{High-quality tailored-edge cleaving using
  aberration-corrected bessel-like beams}.
\newblock \emph{\bibinfo{journal}{Opt. Lett.}} \textbf{\bibinfo{volume}{43}},
  \bibinfo{pages}{3164} (\bibinfo{year}{2018}).

\bibitem{Meyer2019}
\bibinfo{author}{Meyer, R.} \emph{et~al.}
\newblock \bibinfo{title}{Extremely high-aspect-ratio ultrafast {Bessel} beam
  generation and stealth dicing of multi-millimeter thick glass}.
\newblock \emph{\bibinfo{journal}{Appl. Phys. Lett.}}
  \textbf{\bibinfo{volume}{114}}, \bibinfo{pages}{201105}
  (\bibinfo{year}{2019}).

\bibitem{Durnin1987}
\bibinfo{author}{Durnin, J.}, \bibinfo{author}{Miceli, J.~J.} \&
  \bibinfo{author}{Eberly, J.~H.}
\newblock \bibinfo{title}{Diffraction-free beams}.
\newblock \emph{\bibinfo{journal}{Phys. Rev. Lett.}}
  \textbf{\bibinfo{volume}{58}}, \bibinfo{pages}{1499--1501}
  (\bibinfo{year}{1987}).

\bibitem{Alexeev2010}
\bibinfo{author}{Alexeev, I.}, \bibinfo{author}{Leitz, K.~H.},
  \bibinfo{author}{Otto, A.} \& \bibinfo{author}{Schmidt, M.}
\newblock \bibinfo{title}{{Application of Bessel beams for ultrafast laser
  volume structuring of non transparent media}}.
\newblock \emph{\bibinfo{journal}{Physics Procedia}}
  \textbf{\bibinfo{volume}{5}}, \bibinfo{pages}{533--540}
  (\bibinfo{year}{2010}).

\bibitem{Boucher2018}
\bibinfo{author}{Boucher, P.} \emph{et~al.}
\newblock \bibinfo{title}{{Generation of high conical angle {Bessel–Gauss}
  beams with reflective axicons}}.
\newblock \emph{\bibinfo{journal}{Appl. Opt.}} \textbf{\bibinfo{volume}{57}},
  \bibinfo{pages}{6725} (\bibinfo{year}{2018}).

\bibitem{Glezer1997}
\bibinfo{author}{Glezer, E.~N.} \& \bibinfo{author}{Mazur, E.}
\newblock \bibinfo{title}{Ultrafast-laser driven micro-explosions in
  transparent materials}.
\newblock \emph{\bibinfo{journal}{Appl. Phys. Lett.}}
  \textbf{\bibinfo{volume}{71}}, \bibinfo{pages}{882} (\bibinfo{year}{1997}).

\bibitem{Vailionis2011}
\bibinfo{author}{Vailionis, A.} \emph{et~al.}
\newblock \bibinfo{title}{Evidence of superdense aluminium synthesized by
  ultrafast microexplosion}.
\newblock \emph{\bibinfo{journal}{Nat. Commun.}} \textbf{\bibinfo{volume}{2}},
  \bibinfo{pages}{445} (\bibinfo{year}{2011}).

\bibitem{Bhuyan2017}
\bibinfo{author}{Bhuyan, M.~K.} \emph{et~al.}
\newblock \bibinfo{title}{Ultrafast laser nanostructuring in bulk silica, a
  {\textquotedblleft}slow{\textquotedblright} microexplosion}.
\newblock \emph{\bibinfo{journal}{Optica}} \textbf{\bibinfo{volume}{4}},
  \bibinfo{pages}{951} (\bibinfo{year}{2017}).

\bibitem{Stoian2002}
\bibinfo{author}{Stoian, R.} \emph{et~al.}
\newblock \bibinfo{title}{{Laser ablation of dielectrics with temporally shaped
  femtosecond pulses}}.
\newblock \emph{\bibinfo{journal}{Appl. Phys. Lett.}}
  \textbf{\bibinfo{volume}{80}}, \bibinfo{pages}{353--355}
  (\bibinfo{year}{2002}).

\bibitem{Semerok2004}
\bibinfo{author}{Semerok, A.} \& \bibinfo{author}{Dutouquet, C.}
\newblock \bibinfo{title}{Ultrashort double pulse laser ablation of metals}.
\newblock \emph{\bibinfo{journal}{Thin Solid Films}}
  \textbf{\bibinfo{volume}{453-454}}, \bibinfo{pages}{501--505}
  (\bibinfo{year}{2004}).

\bibitem{Chowdhury2005}
\bibinfo{author}{Chowdhury, I.~H.}, \bibinfo{author}{Xu, X.} \&
  \bibinfo{author}{Weiner, A.~M.}
\newblock \bibinfo{title}{Ultrafast double-pulse ablation of fused silica}.
\newblock \emph{\bibinfo{journal}{Appl. Phys. Lett.}}
  \textbf{\bibinfo{volume}{86}}, \bibinfo{pages}{151110}
  (\bibinfo{year}{2005}).

\bibitem{Cao2018}
\bibinfo{author}{Cao, Z.} \emph{et~al.}
\newblock \bibinfo{title}{Influence of electron dynamics on the enhancement of
  double-pulse femtosecond laser-induced breakdown spectroscopy of fused
  silica}.
\newblock \emph{\bibinfo{journal}{Spectrochim. Acta, Part B}}
  \textbf{\bibinfo{volume}{141}}, \bibinfo{pages}{63--69}
  (\bibinfo{year}{2018}).

\bibitem{Gaudfrin2020}
\bibinfo{author}{Gaudfrin, K.} \emph{et~al.}
\newblock \bibinfo{title}{Fused silica ablation by double femtosecond laser
  pulses: influence of polarization state}.
\newblock \emph{\bibinfo{journal}{Opt. Express}} \textbf{\bibinfo{volume}{28}},
  \bibinfo{pages}{15189} (\bibinfo{year}{2020}).

\bibitem{Nagata2005}
\bibinfo{author}{Nagata, T.}, \bibinfo{author}{Kamata, M.} \&
  \bibinfo{author}{Obara, M.}
\newblock \bibinfo{title}{Optical waveguide fabrication with double pulse
  femtosecond lasers}.
\newblock \emph{\bibinfo{journal}{Appl. Phys. Lett.}}
  \textbf{\bibinfo{volume}{86}}, \bibinfo{pages}{251103}
  (\bibinfo{year}{2005}).

\bibitem{Wortmann2007}
\bibinfo{author}{Wortmann, D.}, \bibinfo{author}{Ramme, M.} \&
  \bibinfo{author}{Gottmann, J.}
\newblock \bibinfo{title}{Refractive index modification using fs-laser double
  pulses}.
\newblock \emph{\bibinfo{journal}{Opt. Express}} \textbf{\bibinfo{volume}{15}},
  \bibinfo{pages}{10149} (\bibinfo{year}{2007}).

\bibitem{Chu2017}
\bibinfo{author}{Chu, D.} \emph{et~al.}
\newblock \bibinfo{title}{Effect of double-pulse-laser polarization and time
  delay on laser-assisted etching of fused silica}.
\newblock \emph{\bibinfo{journal}{J. Phys. D: Appl. Phys.}}
  \textbf{\bibinfo{volume}{50}}, \bibinfo{pages}{465306}
  (\bibinfo{year}{2017}).

\bibitem{Wang2017}
\bibinfo{author}{Wang, Z.} \emph{et~al.}
\newblock \bibinfo{title}{High-throughput microchannel fabrication in fused
  silica by temporally shaped femtosecond laser {Bessel}-beam-assisted chemical
  etching}.
\newblock \emph{\bibinfo{journal}{Opt. Lett.}} \textbf{\bibinfo{volume}{43}},
  \bibinfo{pages}{98} (\bibinfo{year}{2017}).

\bibitem{Wang2018}
\bibinfo{author}{Wang, H.}, \bibinfo{author}{Song, J.}, \bibinfo{author}{Li,
  Q.}, \bibinfo{author}{Zeng, X.} \& \bibinfo{author}{Dai, Y.}
\newblock \bibinfo{title}{Formation of nanograting in fused silica by
  temporally delayed femtosecond double-pulse irradiation}.
\newblock \emph{\bibinfo{journal}{J. Phys. D: Appl. Phys.}}
  \textbf{\bibinfo{volume}{51}}, \bibinfo{pages}{155101}
  (\bibinfo{year}{2018}).

\bibitem{Stankevic2020}
\bibinfo{author}{Stankevi{\v{c}}, V.}, \bibinfo{author}{Karosas, J.},
  \bibinfo{author}{Ra{\v{c}}iukaitis, G.} \& \bibinfo{author}{Ge{\v{c}}ys, P.}
\newblock \bibinfo{title}{Improvement of etching anisotropy in fused silica by
  double-pulse fabrication}.
\newblock \emph{\bibinfo{journal}{Micromachines}}
  \textbf{\bibinfo{volume}{11}}, \bibinfo{pages}{483} (\bibinfo{year}{2020}).

\bibitem{Kerse2016}
\bibinfo{author}{Kerse, C.} \emph{et~al.}
\newblock \bibinfo{title}{Ablation-cooled material removal with ultrafast
  bursts of pulses}.
\newblock \emph{\bibinfo{journal}{Nature}} \textbf{\bibinfo{volume}{537}},
  \bibinfo{pages}{84--88} (\bibinfo{year}{2016}).

\bibitem{Bonitz2020}
\bibinfo{author}{Bonitz, M.} \emph{et~al.}
\newblock \bibinfo{title}{Ab initio simulation of warm dense matter}.
\newblock \emph{\bibinfo{journal}{Phys. Plasmas}}
  \textbf{\bibinfo{volume}{27}}, \bibinfo{pages}{042710}
  (\bibinfo{year}{2020}).

\bibitem{Froehly2014}
\bibinfo{author}{Froehly, L.}, \bibinfo{author}{Jacquot, M.},
  \bibinfo{author}{Lacourt, P.~A.}, \bibinfo{author}{Dudley, J.~M.} \&
  \bibinfo{author}{Courvoisier, F.}
\newblock \bibinfo{title}{{Spatiotemporal structure of femtosecond Bessel beams
  from spatial light modulators}}.
\newblock \emph{\bibinfo{journal}{J. Opt. Soc. Am. A}}
  \textbf{\bibinfo{volume}{31}}, \bibinfo{pages}{790} (\bibinfo{year}{2014}).

\bibitem{Rapp2016}
\bibinfo{author}{Rapp, L.} \emph{et~al.}
\newblock \bibinfo{title}{High aspect ratio micro-explosions in the bulk of
  sapphire generated by femtosecond {Bessel} beams}.
\newblock \emph{\bibinfo{journal}{Sci. Rep.}} \textbf{\bibinfo{volume}{6}},
  \bibinfo{pages}{34286} (\bibinfo{year}{2016}).

\bibitem{Dudutis2018}
\bibinfo{author}{Dudutis, J.}, \bibinfo{author}{Stonys, R.},
  \bibinfo{author}{Ra{\v{c}}iukaitis, G.} \& \bibinfo{author}{Ge{\v{c}}ys, P.}
\newblock \bibinfo{title}{Aberration-controlled {Bessel} beam processing of
  glass}.
\newblock \emph{\bibinfo{journal}{Opt. Express}} \textbf{\bibinfo{volume}{26}},
  \bibinfo{pages}{3627} (\bibinfo{year}{2018}).

\bibitem{Goette2016}
\bibinfo{author}{Götte, N.} \emph{et~al.}
\newblock \bibinfo{title}{Temporal airy pulses for controlled high aspect ratio
  nanomachining of dielectrics}.
\newblock \emph{\bibinfo{journal}{Optica}} \textbf{\bibinfo{volume}{3}},
  \bibinfo{pages}{389} (\bibinfo{year}{2016}).

\bibitem{Lin2010}
\bibinfo{author}{Lin, C.-H.} \emph{et~al.}
\newblock \bibinfo{title}{Investigations of femtosecond{\textendash}nanosecond
  dual-beam laser ablation of dielectrics}.
\newblock \emph{\bibinfo{journal}{Opt. Lett.}} \textbf{\bibinfo{volume}{35}},
  \bibinfo{pages}{2490} (\bibinfo{year}{2010}).

\bibitem{Ito2018}
\bibinfo{author}{Ito, Y.}, \bibinfo{author}{Yoshizaki, R.},
  \bibinfo{author}{Miyamoto, N.} \& \bibinfo{author}{Sugita, N.}
\newblock \bibinfo{title}{Ultrafast and precision drilling of glass by
  selective absorption of fiber-laser pulse into femtosecond-laser-induced
  filament}.
\newblock \emph{\bibinfo{journal}{Appl. Phys. Lett.}}
  \textbf{\bibinfo{volume}{113}}, \bibinfo{pages}{061101}
  (\bibinfo{year}{2018}).

\bibitem{Grojo2010}
\bibinfo{author}{Grojo, D.} \emph{et~al.}
\newblock \bibinfo{title}{Exciton-seeded multiphoton ionization in bulk
  {SiO}2}.
\newblock \emph{\bibinfo{journal}{Phys. Rev. B}} \textbf{\bibinfo{volume}{81}}
  (\bibinfo{year}{2010}).

\bibitem{Garcia-Lechuga2017}
\bibinfo{author}{Garcia-Lechuga, M.} \emph{et~al.}
\newblock \bibinfo{title}{Simultaneous time-space resolved reflectivity and
  interferometric measurements of dielectrics excited with femtosecond laser
  pulses}.
\newblock \emph{\bibinfo{journal}{Phys. Rev. B}} \textbf{\bibinfo{volume}{95}}
  (\bibinfo{year}{2017}).

\bibitem{Engelhorn2015}
\bibinfo{author}{Engelhorn, K.} \emph{et~al.}
\newblock \bibinfo{title}{Electronic structure of warm dense silicon dioxide}.
\newblock \emph{\bibinfo{journal}{Phys. Rev. B}} \textbf{\bibinfo{volume}{91}}
  (\bibinfo{year}{2015}).

\bibitem{Denoeud2014}
\bibinfo{author}{Denoeud, A.} \emph{et~al.}
\newblock \bibinfo{title}{Metallization of warm dense {SiO}2 studied by {XANES}
  spectroscopy}.
\newblock \emph{\bibinfo{journal}{Phys. Rev. Lett.}}
  \textbf{\bibinfo{volume}{113}} (\bibinfo{year}{2014}).

\end{thebibliography}

\end{document}